# An Intuitive and Exact Steady-State Electrodynamic Formalism for Uniaxial Multilayered Structures: Normal Incidence

Pieder Beeli[*]
Texas Center for Superconductivity and Advanced Materials; University of Houston; Houston, TX 77204-5002

Date received__________

Exact expressions for all the steady-state fields (*E, H, D, B*) in uniaxial linear media composed of an arbitrary number of layers having arbitrary thicknesses subjected to normal incidence are derived. Generic boundary condition relations in terms of the surface wave impedance are applied at each surface so that fields between any sequence of layers can be related by a cascaded multiplication of transfer functions. With the substitution of the appropriate surface wave impedance for the generic surface wave impedance, these generic transfer functions can be made to represent any reflection or transmission of a wave across any boundary. This formalism obviates the need to solve a large set of equations or an infinite series of reflections and transmissions, which has been the traditional approach in solving such problems. A numerically robust exact expression for the power dissipating in any uniaxial layer is also provided. Examples of the analysis of multilayered systems are given. Although the development is devoted to electromagnetic waves, the methodology and expressions are transferable to acoustical waves, and to some extent, the quantum mechanical wave function.

## 1. INTRODUCTION

The electrodynamics of multilayered materials plays an important role in providing quantitative analysis for basic materials research. Thin film technology underlies the greatest scientific and technological advances of the previous century, including the transistor, VLSI and optical components. This broad panoply of basic research potential and technological demand adds to the need to develop an accurate and intuitive electrodynamic formalism.

Although the optics community has long used exact methods to analyze multilayered systems[1] the complexity of this formalism, its detachment from intuitive physical meaning and the absence of analytic expressions for the fields and power dissipation in each layer compelled workers in the materials science community to seek more simple approaches. Such approaches can be seen in the superconductivity literature which considers electrodynamic transmission through a superconducting thin film and an adjacent dielectric on which the film was deposited. The analysis for these early transmission experiments ignored the dissipation in the dielectric substrate, assumed the *E*-field to be constant throughout the film, ignored the magnetic flux in the film and took the substrate thickness to be infinite.[2,3,4,5,6] This latter approximation is particularly costly because it fails to capture the important effect of the standing waves in the substrate. To account for this difficulty Tinkham and Ginsberg argue that to within 0.2 % the "averaging" done by having a *band* of frequencies impinge on this film-plus-substrate structure yields the same transmissivity as the single frequency impinging on a film plus semi-infinite substrate [4]. However that analysis uses an "exact" transmissivity expression that is independent of the loss tangent of the dielectric substrate and, inter alia, takes as a simplifying assumption a frequency band which linearly increases and then decreases in intensity with frequency. Forthcoming work shows that with the simplifying assumption of constant intensity over the band, errors over 20 % result.[7] More recently Ceremuga-Mazierska[8] follows these approximations except with a refinement made by transforming impedances via the exact transmission line formalism so that the thickness of the dielectric substrate is accounted. However, as discussed in sec. 4C, other problems with [8] arise.

General electrodynamic text books fail to provide a lucid formalism for thin-film electrodynamics. Ramo, Winnery and Van Duzer[9] provide one of the more elucidating treatments, showing that the ratio of field amplitudes--evaluated at the surface of the film--determines the surface wave impedance. However these authors obfuscate the simple underlying physics by their choice of basis function to represent the fields in the film. They choose a hyperbolic basis function instead of a simple exponential function which has a more simple solution and a straightforward interpretation in terms of a steady-state forward and backward propagating wave.[10] Stratton[11] considers the case of normal incidence at a film and develops transfer functions between the incident, reflected and transmitted field (across a film), but the expressions are not generalizable to media consisting of multiple layers. Due to the complexity and unintuitive character of the formalism employed by Stratton, he incurs an error in the transmissivity expression which is noted elsewhere.[7] Stratton's relations do not explicitly use the wave impedance at a surface, but instead explicitly use only the intrinsic impedances. The use of the surface wave impedance is a key feature that ushers the simplification and elucidation contained herein.

An honest treatment of multilayered electrodynamics is even difficult to find in general graduate-level electrodynamics texts which tend to fixate on semi-infinite media. Jackson writes, "This chapter is concerned with plane waves in unbounded, or perhaps semiinfinite, media," but then introduces a problem involving normal incidence at a metal "having a thickness D."[12] The solution shown for this problem is rather unintuitive and not generalizable to multilayered systems. Eqns. 13, 14, 17, 20 and 22 herein reveal that fundamentally only two factors determine the fields and the dissipation: *the impedance mismatch between the wave impedance at the surface of a film and the intrinsic impedances on either side of the surface* and *the intralayer propagation of the fields through the layers*. Yet "impedance mismatch" does not merit a listing in the index of [12]. This latter criticism also applies to the text by Landau and Lifshitz[13] which claims that for the case of "an absorbing medium…. the explicit expressions for the amplitude and phase relations between the three waves [*i.e.*, the incident, reflected and refracted waves] are then extremely involved; they are given by [Stratton, ref. 11 Chapter IX]." To wit, for the more general case of oblique incidence, Stratton derives equations only after untold "laborious calculation[s]" and then says, "The complexity of what appeared at first to be the simplest of problems--the reflection of a plane wave from a plane, absorbing surface--is truly amazing" [11]. In contrast--for the case of normal incidence, but for arbitrary uniaxial dissipative or loss-less materials--we show that these relations can have a simple and exact form that can be written by inspection. As will be shown in a forthcoming paper, the case of normal incidence can be used to guide the more general case of oblique incidence. The forthcoming oblique formalism produces exact and intuitive expressions which can also be produced with a few simple transfer functions.

There are specialty texts which consider stratified media. The text by Wait[14] makes many beautiful contributions and nicely structures the more general problem of oblique incidence. While he provides recursion relations for the wave impedance, he fails to develop these corresponding recursion relations for the fields, but instead directs the reader to the "2(*M-1*) equations which are linear in [the field amplitudes] to solve for 2(*M*-1) unknowns in terms of the [incident field amplitude]." Wait's emphasis is on remote sensing rather than basic materials research and this makes the sub-surface physics less essential for him. An important text by Brekhovskikh[15] begins to provide an approach that is similar to that taken herein in that a recursive transfer function for acoustic pressure fields is provided in Eq. 3.47 of sec. 3.6 (equivalent to our Eq. 11a). Eq. 3.42 nicely describes the acoustic pressure in each layer, but the subsequent discussion on electromagnetics does not employ this elegant transfer function approach as promised. Although Brekhovskikh rightly argues that "the same mathematical methods may be applied" to either acoustic or electromagnetic phenomena, he fails to provide electromagnetic expressions which employ the simplifying transfer function formalism. After deriving this field recursion relation, this sec. 3.6 declares, "The theory developed above has wide application, to which Chapter II is especially devoted." Neither in Chapter II nor anywhere else does the text illustrate the utility of this useful recursion relation to determine the electromagnetic fields or the power dissipation. Instead, this field recursion relation--which involves surface wave impedances--is abandoned. Also abandoned is the surface wave impedance in favor of intrinsic impedances with their complicated entourage of exponentials and multiplied terms. This only complicates the formalism and makes it rather identical to previous work. Brehkhovskikh doesn't exploit this field transfer function across a layer to lucidly reveal other relationships but provides complicated expressions where simpler and more elucidating ones are available [16]. In contrast, the use of transfer functions elegantly abandons the *N* by *N* system of equations approach and enables one to solve the problem by inspection. By leveraging off the simplicity and power of transfer functions, the epoch making paper by Parratt[17] produced a transfer function recursion relation for the reflectivity for oblique incidence and transformed the discipline of X-ray depth profile characterization of multilayered structures. This work expands the usage of transfer functions and shows that by means of a few generic transfer functions *any* electrodynamic transfer function can be developed with ease and insight. It is hoped that the formalism presented herein would spawn an analogous result in materials characterization.

While the results are usually verisimiltudinous, the optics literature expresses reflectances and transmittances in very complicated and lengthy expressions. Heavens[18] shows long and complicated expressions for the reflection coefficient. This complexity makes a strong case for the simple formulas developed herein, which enable one to exactly answer any such question (transfer function for power or phase or magnitude of any fields) in a form that can basically be written by inspection. Likewise Crook[19] provides complicated expressions based on the temporal series approach, which involves taking an infinite sum of all reflections. Using an admittance instead of an impedance formalism, Salzberg[20] shows general oblique relations expressing the fields on one side of a slab to the other, but does not use these relations to determine the fields in an arbitrary layer, the energy dissipated in a layer or to develop any sort of cascading of these transfer functions to produce another transfer function. P. Yeh[21] provides matrix methods by which one can compute the fields in each medium, but does not provide explicit relations for the fields in each media. Instead of only using intrinsic impedances in the reflection/transmission coefficients, we use the surface wave impedance. All the results of secs. 2 and 3 herein include all reflections and transmissions appropriate to each surface and yield the exact steady-state result. This is done without employing an infinite series that arises when one does not use the surface wave impedance. The surface wave impedance by its nature accounts for the infinite temporal series of reflections and can be computed exactly and recursively. What distinguishes this work from other work is the use of surface wave impedances to generate exact field and power expressions. We do so by solving a general boundary condition once and then applying this relationship to each boundary. The surface wave impedance of each surface is found by using the intrinsic impedance of the final layer and using an impedance transformation in a backward recursive scheme, until the wave impedance at the first surface is obtained. Using these impedances now determined at each surface, the field transfer functions at each surface can be generated. Then, by working from the front surface and recursively moving through each surface until arriving at the final (back) surface, one can determine the fields throughout the multilayered structure in terms of the incident field. This intuitive approach is considerably simpler than solving $N$ equations in $N$ unknowns which produces intractable expressions that obfuscate physical intuition.[11,14,18,19,22]

Recently, an exact canonical formalism for the power dissipating in an $N$-layered structure with arbitrary thicknesses was provided.[23] As is the case with this work, the only qualification on the thickness of the films is that the length scales being considered be sufficiently large to average free electrons with donor ions so that $\nabla \cdot \boldsymbol{E}$ can be ignored. We model the conduction as local and linear--governed by Ohm's Law [24]. Similarly, the constitutive relations are taken to be linear. As shown in sec. 2, the satisfaction of these three approximations yields the Helmholtz equation. Working under these general assumptions, Ref. [23] corrects long-standing putative notions of the wave impedance in thin materials satisfying the good conductor approximation and more recent claims about the effect of dielectrics in structures involving thin films. However, because [23] is only concerned with the steady-state *power* dissipation, it ignores the phase of the fields and does not provide explicit field expressions. This work gives the exact expressions for the steady-state fields in each layer of such an $N$-layered structure (Fig. 1) and provides numerically superior algorithms for computing the power in an arbitrary layer.

Although the claim is sometimes made that the solution for the expressions of all the fields in a single film of arbitrary thickness in the good conductor approximation has yet to be published,[25] in fact the solution does exist in the optics literature [21]. However the solutions familiar to the Optics community are mathematically cumbersome. Matrixes are used and the physical essence can be obtuse. Workers look for solutions that can be "simplified considerably" at least for a range of variables (like the thin film limit, the good conductor approximation, …).[26] This work marks the first appearance of these expressions for a single film--or an arbitrary number of films--in this intuitive and simple form. Unfortunately, workers sometimes over simplify. Ghosh *et al.*[27] claim to "investigate the microwave losses of YBCO thin films" but only offer *bulk* expressions. Similarly Lorrain and Corson[28] pedagogically progress from bulk systems to thin films, but then use the bulk transfer function to relate the incident field to the field inside the thin film. But a thin film field requires a thin film transfer function. N.-C. Yeh *et al.*[29] purport to do analysis applicable for films "with a *general thickness*," but do not impose a single boundary condition in their analysis.

Approximate results which hold in certain limits have been forwarded. For example Fahy, Kittel and Louie provide a simple expression for the power transmitted, reflected and dissipated in a thin metal film [26]. They specify thickness limitations and make the good conductor approximation. As will be shown in sec. 4A, their relations encounter errors approaching 100% even when the system is within this parameter space. Although they specify the good conductor approximation, they do not explicitly specify other conductivity limitations. As will be seen in sec. 4B their relations do not hold well for a general complex conductivity, even if this complex conductivity satisfies the good conductor approximation.

Of central importance is the wave impedance and so one is especially interested in simplified approximate thin-film surface wave impedance ($Z_{sw}$) expressions, which evaluate the ratio of $E$ to $H$ at the surface of a structure. Often the sheet impedance result ($Z_{sh} \equiv 1/\sigma d$)--which stems from the dc formula for $V/I$ for a square film of thickness $d$--is substituted for $Z_{sw}$ for frequencies as high as the microwave region [9] or even the far-infrared [2]. Gittleman and Rosenblum[30] glibly write, "Since the film is thin, $j$ [the current density] is uniform in it, and there is really no boundary problem to solve." Although the sheet impedance formula was shown long ago to be a valid approximation for $Z_{sw}$ for a metal film on a bulk metal provided certain unspecified conditions were satisfied,[31] $Z_{sh}$ is not applicable in general for ac transport in conductive films, nor have the conditions for its applicability been precisely delineated. Recent work has determined that there are three necessary conditions that must be satisfied by a metallic film in order for $Z_{sh}$ to have applicability in the case of bimetallic structures.[32] These conditions involve both the film thickness and the impedance mismatch between the two metals. Nonetheless workers continue to make inappropriate or indiscriminate use of the sheet impedance formula. The purpose of the formalism of sec. 2 is to provide exact and simple results which utilize physically intuitive transfer functions. Further simplification can be achieved due to the modular structure of the formalism whereby wave impedance approximations can be immediately inserted to directly translate wave impedance simplification to field transfer function simplification.

## 2. DEVELOPMENT OF THE TRANSFER FUNCTION FORMALISM

We begin by defining for some medium $i$, $\gamma_i \equiv \alpha_i + j\beta_i$ to be the propagation constant where $\alpha_i$ and $\beta_i$ are real, $j \equiv \sqrt{-1}$ and $\alpha_i \equiv 1/\delta_{Ai}$ and $\beta_i \equiv 2\pi/\lambda_i \equiv 1/\delta_{Pi}$ where $\delta_{A(P)i}$ is the amplitude attenuation(phase) length scale and $\lambda_i$ is the wavelength in medium $i$. The temporal dependence is taken to be $e^{j\omega t}$ so that $\delta_{Ai}$ and $\beta_i$ are necessarily non-negative.[33] This convention yields a bulk spatial-temporal form of $\exp(-\gamma x + j\omega t)$. Since we limit ourselves to linear media, it suffices to determine $E$ and $H$, from which $D$ and $B$ are immediately determined by the constitutive equations: $D_i = \varepsilon_i E_i$ and $B_i = \mu_i H_i$ for $i = x, y, z$.

As shown in Fig. 1, physically the problem we undertake is three dimensional. However because there is no transverse structure, for mathematical purposes the problem can be reduced to one dimension, which we denote by *x*. Each point on the *x* axis corresponds to a plane in *y-z* space. This plane, or surface, gives rise to the association of a surface with a point and to our most central descriptor, the surface wave impedance.

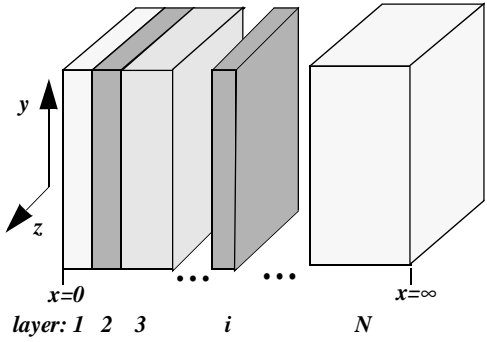

FIG. 1 Geometry of the *N*-layer structure under consideration. The incident wave emerges from medium 0 and impinges normally upon the surface at $x = 0$. Each layer has its uniaxis in the ***x*** direction--which is parallel to the direction of the incident radiation. Otherwise, each layer can be of arbitrary composition (*E.g.,* a metal, insulator, superconductor, semiconductor …) and of rather arbitrary thickness.

In order to have meaningful power dissipation relations we assume that medium 0, the medium in which the fields originate, has an intrinsic wave impedance that is purely real so that medium 0 is lossless.[34]

### A. Homogeneous Helmholtz equation in *H*

From Maxwell's modification of Ampere's Law,

$$\nabla \times \vec{H} = \vec{J} + \frac{\partial}{\partial t}\vec{D}, \qquad (1)$$

we assume Ohm's law [24], the steady-state and linear media to obtain

$$\nabla \times \vec{H} = \Sigma \otimes \vec{E} + j\omega\varsigma \otimes \vec{E} \qquad (2)$$

where the $\otimes$ symbol denotes multiplication of the conductivity tensor $\Sigma$ (which we take to be diagonal with elements $\sigma_c$, $\sigma_{a\text{-}b}$ and $\sigma_{a\text{-}b}$ to be equal to $\sigma_{xx}$, $\sigma_{yy}$ and $\sigma_{zz}$ respectively) and permittivity tensor $\varsigma$ (with diagonal elements equal to $\varepsilon_{xx}$, $\varepsilon_{a\text{-}b}$ and $\varepsilon_{a\text{-}b}$). Taking the curl of Eq. 2, we obtain[35]

$$\nabla(\nabla \cdot \vec{H}) - \nabla^2\vec{H} = \nabla \times j\omega\varsigma_c \otimes \vec{E} \qquad (3)$$

where we have defined the complex permittivity tensor ($\varsigma_c$) via $j\omega\varsigma_c \equiv \Sigma + j\omega\varsigma$.

Next we confine our geometry as per Fig. 1 where the transverse spatial derivatives are zero. After utilizing the absence of the magnetic monopole and confining *E* and *H* to be transverse, we obtain

$$-\nabla^2(0, H_y, H_z) = j\omega\varepsilon_{c(a\text{-}b)}\left(0, -\frac{\partial}{\partial x}E_z, \frac{\partial}{\partial x}E_y\right) \qquad (4a)$$

where $\varepsilon_{c(a\text{-}b)}$ is the complex transverse permittivity tensor element, equal to $\varepsilon_{a\text{-}b}(1 - j\sigma_{a\text{-}b}/\omega\varepsilon_{a\text{-}b})$. But from Lenz's Law ($\nabla \times \vec{E} = -\partial\vec{B}/\partial t$) we obtain

$$\left(0, -\frac{\partial}{\partial x}E_z, \frac{\partial}{\partial x}E_y\right) = -\partial\vec{B}/\partial t = -j\omega\mathbf{M} \otimes \vec{H} \qquad (4b)$$

where $\mathbf{M}$ is the diagonal permeability tensor whose elements are $\mu_{xx}$, $\mu_{a\text{-}b}$ and $\mu_{a\text{-}b}$. Inserting Eq. 4b into Eq. 4a we obtain, the homogenous Helmholtz equation in $H_y$ and $H_z$:

$$\nabla^2(0, H_y, H_z) = (j\omega)^2\varepsilon_{c(a\text{-}b)}\mu_{a\text{-}b}(0, H_y, H_z)$$

$$\equiv \gamma^2(0, H_y, H_z) \qquad (5)$$

where we have thus shown that $\gamma$, the propagation constant, is given by $j\omega\sqrt{\mu_{a\text{-}b}\varepsilon_{c(a\text{-}b)}}$. Eq. 5 reveals that the *transverse* fields are independent of the *longitudinal* tensor elements, and that the *longitudinal* propagation constant is independent of the *longitudinal* tensor elements.

Since we can always rotate the uniaxial system of Fig. 1 about the *x*-axis, it is always possible to align the left incident *E*-field with the *y*-axis and to take the corresponding *H*-field to be in the *z*-direction. Subsequently we may omit these superfluous orientational subscript designations on these tensor elements. A similar development reveals that Eq. 5 is analogously obeyed by $\vec{E}$.

We represent the solution to the Helmholtz equation of the *H*-field in the $i^{th}$ layer by both a forward ($A_i$) and backward rms amplitude ($B_i$). We follow the notation in [23] and represent the location of the boundary separating the *i-1*$^{th}$ layer from the $i^{th}$ layer by $x_i$. Our development begins with the determination of the *H*-field, and from this field the *E*-field is determined via Ampere's Law. In phasor notation, we take the general solution to Eq. 5 to be

$$H_i(x, t) = e^{j\omega t}(A_i e^{-\alpha_i\Delta x_i - j\beta_i\Delta x_i} + B_i e^{\alpha_i\Delta x_i + j\beta_i\Delta x_i}) \qquad (6a)$$

for *i*=0…*N*-1 where $0 \leq \Delta x_i \equiv x - x_i \leq d_i \equiv x_{i+1} - x_i$ where $d_i$ is the thickness of the $i^{th}$ layer so that only *x* satisfying $x_i \leq x \leq x_{i+1}$ are applicable for $H_i(x)$. Further, $\gamma_i \equiv \alpha_i + j\beta_i = j\omega\sqrt{\mu_i\varepsilon_i(1 - j\sigma_i/\varepsilon_i\omega)}$ where $\mu_i$, $\varepsilon_i$ and $\sigma_i$ are the transverse permeability, permittivity and conductivity components of medium *i* respectively.[36] $\mu_i$ and $\varepsilon_i$ are taken to be real, while $\sigma_i$ is allowed to be complex. For the $N^{th}$ layer we have

$$H_N(x, t) = e^{j\omega t}A_N e^{-\alpha_N\Delta x_N - j\beta_N\Delta x_N} \qquad (6b)$$

for $x \geq x_N$. In order to be consistent with the taking of a derivative of a continuous function on an atomistic lattice, for all layers composed of condensed matter we require that $\delta_{Ai}$ and $\lambda_i$ be much larger than the lattice spacing in the corresponding layer composed of condensed matter. Because the mean free path is along the transverse direction while the wave propagation is longitudinal, we expect mean free path variations to only have secondary effects.

### B. Surface wave impedance recursion relation

By Maxwell's curl equations, both $E$ and $H$ are continuous over all $x$. Thus it follows that the ratio of $E/H$ will also be continuous. Because each $x$ value defines a surface in the $y$-$z$ plane, this ratio is constant over the surface and is called the surface wave impedance ($Z_{sw}$ or $Z$). In the bulk case, this ratio is independent of the size of the material and is then equal to the intrinsic wave impedance ($\eta$). From $\nabla \times \vec{E} = -\mu \partial \vec{H}/\partial t = -j\omega\mu\vec{H}$, and inserting $\vec{E} = \hat{y}E_y e^{-x\gamma}$, one obtains

$$\eta_i \equiv E_{yi}/H_{zi} = \frac{j\omega\mu_i}{\gamma_i}. \tag{7a}$$

Alternatively, from $\nabla \times \vec{H} = (\sigma + j\omega\varepsilon)\vec{E}$ and Eq. 5 one finds for the $i^{th}$ layer

$$\eta_i \equiv \sqrt{\frac{\mu_i}{\varepsilon_{ci}}} \tag{7b}$$

Although Eq. 7 is derived for the case of a bulk material, it applies for films as well since whether the material be bulk or a thin film, the intrinsic impedance has the same relation to the material parameters. We have therefore put the subscript "$i$" onto the material descriptors of Eq. 7 and allow $i$ to apply to all the media, i.e., $i = 0, 1, 2, \ldots N$.

For the case when there are multiple layers, one is interested in the wave impedance at each surface. The wave impedance at any layer can be found in a recursive fashion beginning with the right-most layer (i.e., the $N^{th}$) and sequentially progressing to the left through each surface until the surface impedance of the incident surface is obtained. The wave impedance at the incident surface of the $i^{th}$ layer ($Z_i$), is given by[33]

$$Z_i = \eta_i(Z_{i+1}M_i + \eta_i N_i)/(\eta_i M_i + Z_{i+1}N_i) \tag{8}$$

where $M_i = \cosh p_i + \cos q_i$, $N_i = \sinh p_i + j\sin q_i$, $p_i = 2\alpha_i d_i$, and $q_i = 2\beta_i d_i$. $Z_N \equiv \eta_N$. One can either take Eqn. 6 (and Eq. 10) as an instantaneous *or* a steady-state relation by using a field transfer function that is determined by intrinsic or surface wave impedances, respectively. Since this work concerns the steady-state response, we employ the latter. The former approach gives rise to the temporal series approach which is a cumbersome way to solve steady-state problems.

Although the wave impedance is generally used properly in bulk work, in thin-film work it is often misapplied. For example one popular effort[37] concerns "the effective microwave surface impedance of high-$t_c$ thin films." This work defines the surface impedance as we define the surface wave impedance, i.e., $E/H$ or $Z_{sw}$. Klein et al.[37] then force a definition on $Z_{sw}$ by imposing the bulk relationship $2 \cdot P_t = R_{sw1}|H_1|^2$ where $P_t$ is the areal power density crossing the incident surface of the film and $R_{sw1}$ is the real part of $Z_{sw1}$. In using this bulk relationship between power and $R_{sw}$, the power is not the power dissipating in the film, but the total power crossing the incident surface of the film. To reconcile this bulk parameter in $R_{sw}$ with the electrodynamics of a thin superconducting film, Klein et al. [37] make an appeal to an "effective" surface resistance which includes the transmitted power beyond the film and into the substrate i.e., "$R_{eff} = R_s f(d/\lambda) + R_{trans}$" where $R_s$ is the real part of the intrinsic wave impedance, $f(d/\lambda)$ is a function that accounts for the finite thickness of the film and $R_{trans}$ is an effective surface wave resistance accounting for the "power transmission into the [assumed semiinfinite] substrate." By appealing to this relation the claim is made that, "$R_{eff}$ is limited to $R_{trans}$ if $R_s$ is zero." Even under the condition $R_s/X_s \ll 1$--for which the expression for $R_{trans}$ is derived--this cannot be generally true. If $R_s$ is zero and we take the limit that the film thickness goes to 0, $R_{trans} \to \infty$ [37]. Clearly this is unrealistic. Rather under these conditions one must find that $R_{eff} \to R_{trans} \to R_{substrate}$ which is approximated by $\eta_o/\sqrt{\varepsilon_r}$.

But more to the point, the use of the term "$R_{trans}$" has questionable utility. One is interested in knowing how much power is dissipating in the film and how much is dissipating in the substrate. Ref. [37] does not tell us how $R_{eff}$ and $R_{trans}$ can answer such questions. The root of the problem lies in simultaneously trying to answer two different questions. One question concerns the thin film and another the thin film system. If one is interested in the dissipation in the film ($P_f$), then the approach of [23] or sec. 3 herein would be appropriate. If one is interested in the total dissipation, then $2 \cdot P_t = R_{sw1}|H_1|^2$ or Eq. 15 can be used, but then what is being measured is no longer the surface resistance of the thin film, but of the thin film system. This effective formalism--even when correct--does not indicate the distribution of the power dissipation, but only that the power is dissipating somewhere in both the film and substrate. Sec. 4C exemplifies the electrodynamic coupling between the layers in a multilayered structure and reveals that the layers form an organic relationship. It is shown there that a small change in one layer can lead to a terrific change in the electrodynamic behavior in another.

### C. Helmholtz equation in E

Having determined the form of $H$ everywhere, we can now proceed to find that of $E$. From $\nabla \times H = J + \partial D/\partial t$, we find $E_i(x,t) = -(\sigma_i + j\omega\varepsilon_i)^{-1}\partial H_i(x,t)/\partial x$. Therefore by Eqns. 5 and 7

$$E_i(x,t) = -\frac{\eta_i}{\gamma_i}\frac{\partial}{\partial x}H_i(x,t). \quad (9)$$

Employing Eq. 9 onto Eq. 6 it follows

$$E_i(x,t) = \eta_i e^{j\omega t}(A_i e^{-\alpha_i \Delta x_i - j\beta_i \Delta x_i} - B_i e^{\alpha_i \Delta x_i + j\beta_i \Delta x_i}) \quad (10a)$$

for $i = 0\ldots N-1$. Analogous to Eq. 6b, we have

$$E_N(x,t) = \eta_N e^{j\omega t} A_N e^{-\alpha_N \Delta x_N - j\beta_N \Delta x_N} \quad (10b)$$

for $x \geq x_N$. Comparing Eq. 6a to Eq. 10a we see that the $H$-field reflection coefficient is of equal magnitude and opposite sign to the $E$-field reflection coefficient.

Defining the field amplitudes at different locations and employing the $\Delta x_i$ in the exponent facilitates an economical description of the fields via a cascading of transfer functions. It is this cascading of transfer functions--as done in sec. 2E--whereby we neatly describe the fields in each region. Because Salzberg [20] does not adopt this $\Delta x_i$ notation, this probably contributes more than any other reason to his omission of expressions for the fields in each medium. Fahy, Kittel and Louie [26] define an $H$-field transmission amplitude at a singular point in real space even though the fields being related in the transfer function do not exist at that point. By defining the transfer function in terms of fields at different locations, we can avoid expressions involving the fields where the fields do not exist.

### D. Field recursion relations

Having determined the fields in terms of complex amplitudes in sec. 2A and 2C, it therefore only remains to determine the complex amplitudes, $A_i$ and $B_i$. We do this by taking the solution to the boundary condition solved in the case of bimetallic systems[10] to be the generic solution for any boundary so that across a boundary dividing layer $i-1$ from layer $i$, we have

$$\frac{A_i(x_i^+)}{A_{i-1}(x_i^-)} = \frac{1 + Z_i/\eta_i}{1 + Z_i/\eta_{i-1}} \quad (11a)$$

and

$$\frac{B_i(x_i^+)}{A_{i-1}(x_i^-)} = \frac{1 - Z_i/\eta_i}{1 + Z_i/\eta_{i-1}} \quad (11b)$$

where $A$ and $B$ have their meaning permuted from that in [10]. Eq. 11 reveals the discontinuity in the field components across a surface whose wave impedance is given by $Z_i$.

From Eqs. 6 and 11 we can determine the field transfer function for the steady-state forward propagating $H$-field rms phasor amplitude from the left surface of the $i$-$1^{th}$ layer ($A_{i-1}$) to the left surface of the $i^{th}$ layer ($A_i$)

$$\frac{A_i}{A_{i-1}} = \frac{1 + Z_i/\eta_i}{1 + Z_i/\eta_{i-1}}e^{-d_{i-1}\gamma_{i-1}} \quad (12a)$$

Similarly we relate the backward propagating steady-state $H$-field amplitude at the left side of the $i^{th}$ layer to the forward propagating steady-state $H$-field amplitude at the left side of the $i$-$1^{th}$ layer

$$\frac{B_i}{A_{i-1}} = \frac{1 - Z_i/\eta_i}{1 + Z_i/\eta_{i-1}}e^{-d_{i-1}\gamma_{i-1}} \quad (12b)$$

for $i = 1\ldots N-1$. Thus the total tangential $H$-field can be found by summing Eqns. 12a and b.

### E. Fields in $i^{th}$ layer

Having determined the field recursion relations, we can now describe the fields in the $i^{th}$ layer in terms of the complex incident forward propagating $H$ field in medium 0 or any other field in any other layer. We begin by establishing the relationship for the forward propagating $H$-field at some arbitrary location to the left of the structure of Fig. 1, $A_0(x,t)$, to the forward propagating $H$-field located just to the left of the structure at the same time, $A_0(0^-,t)$:

$$A_0(0^-,t) = A_0(x,t)e^{-\gamma_0|x|} \quad (13a)$$

where $x \leq 0$ and $\gamma_0$ is, by hypothesis, purely imaginary. Next we express the total field in the originating medium--$H_0(x,t)$--in terms of $A_0(0^-,t)$ for $x \leq 0$

$$\frac{H_0(x,t)}{A_0(0^-,t)} = \left(e^{\gamma_0|x|} + \frac{1 - Z_1/\eta_0}{1 + Z_1/\eta_0}e^{-\gamma_0|x|}\right) \quad (13b)$$

Via Eqns. 13a and 13b we also have the means of expressing the total field for $x \leq 0$ in terms of the forward propagating field for any $x \leq 0$. It remains to compute the fields in each layer in terms of a convenient field amplitude, say $A_0(0^-,t)$. In the first layer, the $H$-field is represented by

$$\frac{H_1(x,t)}{A_0(0^-,t)} = \frac{1}{1 + Z_1/\eta_0}$$

$$\cdot[(1 + Z_1/\eta_1)e^{-\gamma_1 x} + (1 - Z_1/\eta_1)e^{\gamma_1 x}] \quad (13c)$$

where $0 \leq x \leq d_1$ (since $x_1 \equiv 0$, $\Delta x_1 = x$). For subsequent layers,

$$\frac{H_i(x,t)}{A_0(0^-,t)} = \prod_{k=1}^{i-1}\frac{1 + Z_k/\eta_k}{1 + Z_k/\eta_{k-1}}\exp\left\{-\sum_{l=1}^{i-1}\gamma_l d_l\right\}$$

$$\cdot\left(\frac{1 + Z_i/\eta_i}{1 + Z_i/\eta_{i-1}}e^{-\gamma_i \Delta x_i} + \frac{1 - Z_i/\eta_i}{1 + Z_i/\eta_{i-1}}e^{\gamma_i \Delta x_i}\right) \quad (13d)$$

for $i=2\ldots N-1$ where $0 \leq \Delta x_i \leq d_i$. In the last layer,

$$\frac{H_N(x,t)}{A_0(0^-,t)} = \prod_{k=1}^{N}\frac{1+Z_k/\eta_k}{1+Z_k/\eta_{k-1}}\exp\left\{-\sum_{l=1}^{N-1}\gamma_l d_l\right\}e^{-\gamma_N \Delta x_N} \quad (13e)$$

for the field somewhere in the last layer (*i.e.* $0 \leq \Delta x_N$). Additionally, one can express the intralayer field components at one place and time $(x,t)$ to another place and time $(x',t')$ via the relation

$$A_i[B_i](x,t) = A_i[B_i](x',t')e^{j\omega(t-t')\mp\gamma_i(x-x')} \quad (13f)$$

for $i = 0, 1\ldots N$ provided the fields have equilibrated to the steady state for all times between $t'$ and $t$, and where the upper[lower] sign applies to $A_i[B_i]$. Eq. 13f is further constrained: $x_i \leq x, x' \leq x_{i+1}$. By Eq. 6b, $B_N \equiv 0$. $x_0 \equiv -\infty$ and $x_{N+1} \equiv \infty$.

Having determined the $H$-field over all space, we now proceed to determine the $E$-field. We denote the complex amplitude of the forward propagating $E$-field at $x = 0^-$ and time $t$ by $C_0(0^-,t)$. Applying Eq. 9 to Eq. 13 and using $C_0(0^-,t) = \eta_0 A_0(0^-,t)$ the corresponding equations for the electric field are generated

$$C_0(0^-,t) = C_0(x,t)e^{-\gamma_0|x|} \quad (14a)$$

for $x \leq 0$. Next we express the total $E$-field in terms of $C_0(0^-,t)$ for $x \leq 0$

$$\frac{E_0(x,t)}{C_0(0^-,t)} = \left(e^{\gamma_0|x|} - \frac{1-Z_1/\eta_0}{1+Z_1/\eta_0}e^{-\gamma_0|x|}\right) \quad (14b)$$

In the first layer, the $E$-field is represented by

$$\frac{E_1(x,t)}{C_0(0^-,t)} = \frac{\eta_1}{\eta_0}\frac{1}{1+Z_1/\eta_0}$$
$$\cdot[(1+Z_1/\eta_1)e^{-\gamma_1 x} - (1-Z_1/\eta_1)e^{\gamma_1 x}] \quad (14c)$$

where $0 \leq x \leq d_1$. For subsequent layers,

$$\frac{E_i(x,t)}{C_0(0^-,t)} = \frac{\eta_i}{\eta_0}\prod_{k=1}^{i-1}\frac{1+Z_k/\eta_k}{1+Z_k/\eta_{k-1}}\exp\left\{-\sum_{l=1}^{i-1}\gamma_l d_l\right\}$$
$$\cdot\left(\frac{1+Z_i/\eta_i}{1+Z_i/\eta_{i-1}}e^{-\gamma_i \Delta x_i} - \frac{1-Z_i/\eta_i}{1+Z_i/\eta_{i-1}}e^{\gamma_i \Delta x_i}\right) \quad (14d)$$

for $i = 2\ldots N-1$ where $0 \leq \Delta x_i \leq d_i$. In the last layer,

$$\frac{E_N(x,t)}{C_0(0^-,t)} = \frac{\eta_N}{\eta_0}\prod_{k=1}^{N}\frac{1+Z_k/\eta_k}{1+Z_k/\eta_{k-1}}\exp\left\{-\sum_{l=1}^{N-1}\gamma_l d_l\right\}e^{-\gamma_N \Delta x_N} \quad (14e)$$

We also write

$$F(x,t) = F(x,t')e^{j\omega(t-t')} \quad (14f)$$

where $F = H, E, A, B$ or $C$. Taken together, Eqns. 13 and 14 allow us to determine any steady-state field at any point and at any time relative to any other field at any other point or time.

### F. Discussion

Yet waves--and the homogeneous Helmholtz equation (Eq. 5)--have a much broader context than electromagnetism such as acoustics[15] and quantum mechanics, and so the formalism noted herein can be extended to a wide range of Physics. We introduce the concept of impedance as the ratio of two continuous steady-state quantities. The electrodynamic impedance is the ratio of the transverse $E$-field to the $H$-field; the acoustic impedance is the ratio of the pressure to the component of the particle velocity normal to the boundary; the quantum mechanical impedance is the ratio of the wave function to its derivative. For the case of a constant potential energy, the Schrödinger equation reduces to a homogeneous Helmholtz equation,[38] which is the basis for our forward and backward propagating waves. A quantum well with constant potential energy is analogous to a layer in Fig. 1 with constant material parameters.

## 3. POWER DISSIPATION

Important for materials characterization are exact electrodynamic expressions in the linear limit. Empowered with an exact formalism, one can monitor power dissipation, current densities, … in each layer and systematically determine the threshold marking the onset of non-linear behavior. Thermal analysis also requires knowing the spatial distribution of the conductive dissipation.

### A. Bulk material expression

The steady-state fractional power dissipated in a bulk material, or the total power dissipated in the layered structure of Fig. 1, is given by [10]

$$\frac{P_t}{P_{in}} = 4\frac{\eta_o}{|Z_1+\eta_o|^2}\Re(Z_1) \quad (15)$$

where $\Re$ is the real operator. $P_t$ is the steady-state areal power density crossing $x_1$. $P_{in}$ is the areal power density incident upon $x_1$. If the $N^{th}$ layer in Fig. 1 locates the measurement of the transmitted wave, then $P_t$ includes this power too. Ref. [39] mistakenly substitutes $(Z^2 + 2R\eta_o + \eta_o^2)$ for the denominator of Eq. 15. Changing the $Z^2$ term to $|Z|^2$ would rectify this error. Ref. [40] claims Eq. 15 is given by $cR_s/\pi$ (cgs) or $4R_s/\eta_o$ (SI), which is only an approximate result. The same result of Eq. 15 can also be obtained by summing the power dissipated in each layer, via the expressions contained in sec. 3B.

### B. Thin film expressions

Although the power dissipation in the $i^{th}$ layer of such a structure has been determined elsewhere [23], that result involves a difference. In the limit that the layer is very thin, the result--being a small difference between two large numbers--suffers a numerical challenge. We seek an algebraically equivalent, but numerically superior result and begin with

$$P_i = \frac{\Re(\tilde{\sigma}_i)}{2}\int_0^{d_i}|E_i|^2 dx_i \tag{16}$$

where $P_i$ is the areal power density dissipating in the $i^{th}$ layer [33]. Because this integral has been undertaken elsewhere in the electrodynamics of thin films [33], we use that particular solution as a generic solution from which we generate two solutions. The first invokes the good conductor approximation while the second is an exact solution.

But before deriving the expressions for the power dissipated in the individual layers, we write the result for the dissipation in the last layer, the $N^{th}$ layer, which is semi-infinite. We have

$$\frac{P_N}{P_{in}} = \frac{4}{\eta_o} \frac{e^{-\sum_{k=1}^{N-1} p_k}}{\left|1+\frac{\eta_N}{\eta_{N-1}}\right|^2} \prod_{j=1}^{N-1}\left|\frac{1+Z_j/\eta_j}{1+Z_j/\eta_{j-1}}\right|^2 \Re(\eta_N) \tag{17}$$

as noted elsewhere [23]. By substituting good conductor approximation versions for $Z_i$, $\eta_i$ and $p_i$ one obtains a good conductor approximation version of this exact expression. It remains then to compute the dissipation in the previous layers.

#### 1. Numerically robust result for the power dissipated in a thin film in the good conductor approximation

To solve for an expression for the power dissipating in a film where the displacement current is neglected relative to the conduction current, we turn to the solution of Eq. 16 found in [33]. Noting that the meaning of $A$ and $B$ defined in this paper is permuted from that in [33], we can quickly write the result:

$$P_i \approx -|A_i|^2 \mu_i \omega \{\Im(\tilde{\lambda}_i)V_{1i} - \Re(\tilde{\lambda}_i)V_{2i}\} \tag{18}$$

for $i=1...N-1$. $\Im(\Re)$ is the imaginary(real) operator and $\tilde{\lambda}_i \equiv 1/\gamma_i$. Further,

$$V_{1i} = e^{-p_i/2}\sinh\left(\frac{p_i}{2}\right)\left[e^{-p_i}\frac{(1-|L_i|^2)^2 + 4\Im^2(L_i)}{|1+L_i|^4} + 1\right] \tag{19a}$$

and

$$V_{2i} = \frac{e^{-p_i}}{|1+L_i|^2}[2\Im(L_i)(1-\cos q_i) - (1-|L_i|^2)\sin q_i] \tag{19b}$$

where $L_i = Z_{i+1}/\eta_i$ is the impedance mismatch between the surface impedance of the $i+1^{th}$ layer and the intrinsic impedance of the $i^{th}$ layer. $p_i$ and $q_i$ are as per the discussion of Eq. 8. Another helpful abbreviation will prove to be

$$\left|\frac{A_i}{A_0}\right| = \prod_{j=1}^{i}\left|\frac{1+Z_j/\eta_j}{1+Z_j/\eta_{j-1}}\right|\exp\left(-\sum_{k=1}^{i-1}\frac{p_k}{2}\right) \tag{19c}$$

which is immediately found from Eq. 12a. We shall wish to normalize Eq. 18 by the incident power. We express the incident power in terms of the incident $H$-field. Since medium 0 is by hypothesis lossless, $|A_o(x)|$ is independent of position and can therefore be evaluated anywhere in that medium. Since $P_{in} = \eta_o|A_o|^2/2$, we write

$$\frac{P_i}{P_{in}} \approx \frac{2}{\eta_o}\left|\frac{A_i}{A_0}\right|^2(\Re(\eta_i)V_{1i}+\Im(\eta_i)V_{2i}) \tag{20}$$

for $i=1...N-1$ (where we also made use of Eq. 7a). Both $\Im(\eta_i)$ and $\Re(\eta_i)$ are necessarily non-negative for any material [33].

#### 2. Exact and numerically robust result for the power dissipated in a thin film

For materials which do not satisfy the good conductor approximation, an analogous expression to Eq. 20 would be helpful. Following the methodology of [33] in integrating Eq. 16, after some work we write

$$P_i = \frac{\Re(\sigma_i)}{2}|A_i\eta_i|^2(\delta_{Ai}V_{1i}+\delta_{Pi}V_{2i}) \tag{21}$$

where $\delta_{Ai} \equiv 1/\alpha_i$ and $\delta_{Pi} \equiv 1/\beta_i$ which are both non-negative for any medium [33]. We can therefore write

$$\frac{P_i}{P_{in}} = \Re(\sigma_i)\frac{|\eta_i|^2}{\eta_o}\left|\frac{A_i}{A_0}\right|^2(\delta_{Ai}V_{1i}+\delta_{Pi}V_{2i}) \tag{22}$$

Finally we note that a study of sec. 2 and 3 reveals that in order to determine the transfer function for fields between the $i^{th}$ and $j^{th}$ layers (where $j > i$), it is not necessary to know the material parameters to the left of the $i^{th}$ layer. In contrast, the power dissipation involves the fields referenced to the fields in medium 0 and so requires knowledge of each layer.

### 4. EXAMPLES

We consider a few examples of the formalism of secs. 2 and 3, from simple to complex. The first examples concern a free-standing film and the final a 4-layered structure.

#### A. Metallic film suspended in free space

Fahy, Kittel and Louie (FKL) consider a metallic slab suspended in free space and derive some simplified relations.[26]

In their derivation FKL are confronted with the integral of the current density over the thickness of the film and they take the solution to be the conductivity times one-half the sum of the total $E$-field at either side of the film.[26] It is their approximation of this average field that we show next to not be justifiable for arbitrary conductivities. Instead of beginning with the expression

$$<E> \equiv \frac{1}{d}\int_0^d (E^+ e^{-\gamma x} + E^- e^{\gamma x})dx \quad (23a)$$

and making subsequent arguments, they quickly write, "the spatial average electric field in the film is

$$<E> = [E_1(0) + E_3(d)]/2" \quad (23b)$$

where $E_1(0)$ is the total $E$-field at the incident face and $E_3(d)$ is the total $E$-field at the back face. The implicit assumption seems to be that by virtue of $d << \delta$, the magnitude and phase of $E_1(0)$ and $E_3(d)$ must be related by $E_3(d) \approx E_1(d) - (1 + j) \cdot d/\delta$. However this does not follow. Rather from $d << \delta$ it follows that the forward propagating field inside the film at the incident surface (i.e., $\eta_f A_1$ or $E^+$) must not differ significantly in magnitude and phase from the same field at the back surface. Similarly for the backward propagating field in the film. Since the phase of the forward field can be ~180° out of phase with the backward field one cannot conclude that one half the sum of $E_1(0)$ and $E_3(d)$ will equal the average magnitude and phase of the field throughout the film. It is not reasonable to assume that the sum of two complex vectors which vary with $x$ by $\exp(j\kappa x)$ and $\exp(-j\kappa x)$ will generally yield--even for $|\kappa x| << 1$--a resultant whose magnitude and phase are both linear in $x$. If the amplitudes are nearly equal in magnitude and phase then the first term in the series expansions will cancel and the resultant field will vary quadratically across the film. Thus the magnitude and phase of the total field across a film can be non-linear--even when $d << \delta$.

Figure 2 depicts the magnitude of the error in the transmissivity, reflectivity and absorptivity equations of [26] as a function of film thickness, normalized by the skin depth. For illustrative purposes, we take the excitation frequency to be 10 GHz and a resistivity of $10^6$ $\mu\Omega$-$cm$. This corresponds to a skin depth ($\delta$) of ~0.5 mm. The film thickness ($d$) is taken to extend from 3 Å to $2\delta$. One sees that the reflectivity error is almost 100 % for very thin films. This large error in the reflectivity is due to the non-linear current density profile in the film.

A metal has a propagation constant with equal real and imaginary magnitudes so that both the amplitude and phase change on a length scale given by $\delta$. Therefore the wavelength inside the metal is given by $2\pi\delta$. Although the introductory remarks in sec. II of [26] portend simplification for "thickness[es] much less than the skin depth," the derivation of the average electric field in the film makes the looser requirement that "$d << \lambda$" where $\lambda$ is the "vacuum" wavelength. From Fig. 2 one sees that as $d$ increases beyond $\sim\delta/2$ (corresponding to $log_{10}d/\delta \approx -0.3$) the errors quickly get large for the transmissivity, reflectivity and absorptivity. Since the free space wavelength ($\lambda_0$) incident upon a metal can be many orders of magnitude larger than $\delta$, and the since the errors in the transmissivity for a metallic film can exceed 60% for $d < 2\delta$ (Fig. 2), the requirement "$d << \lambda$" is clearly too liberal. This oversight is also made by Ginsberg and Tinkham [4], Tinkham [41] and Sridhar and Mercereau [42].

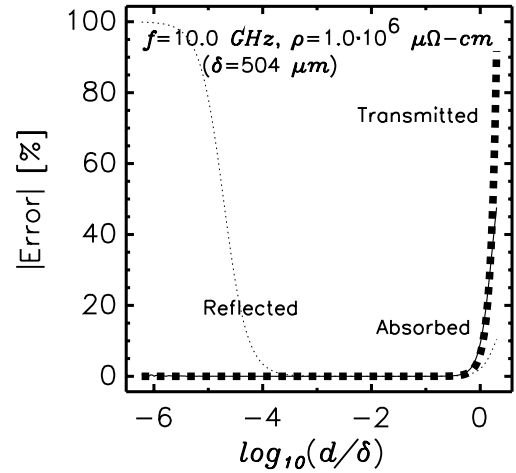

FIG. 2 Error in the transmissivity, reflectivity and absorptivity expressions of [26] for a free standing metallic film as a function of film thickness, normalized by the skin depth ($\delta$). The frequency of the incident radiation is taken to be 10 GHz and the resistivity of the film is $10^6$ $\mu\Omega$-$cm$. The thickness varies from 3 Å to $2\delta$.

The use of Eq. 23b is popular among other authors as well [8]. One advantage of the formalism of sec. 2 is won by not appealing to the approximate approach associated with Eq. 23b.

Nonetheless [26] is one of few sources which discusses the non-monotonic behavior in the absorption as a function of thickness and impedance mismatch. However for many metals--including the Cu example given--the thickness corresponding to the absorption peak is comparable to or less than the nominal lattice spacing. Other work[10,43] illustrates this non-monotonic phenomena as a function of film thicknesses for the full range of thicknesses from 3 Å to $3\delta_A$.

B. Slab of complex conductivity suspended in free space

Eq. 17 can be applied to give the exact result for the transmissivity of a slab whether the slab is composed of a purely real conductivity or conductivity with arbitrary real and imaginary parts. The derivation of a transmissivity, reflectivity and dissipation expression in [26] concerns a purely real conductivity when FKL write "$d_c = c/2\pi\sigma$." For $d_c$ to be real, $\sigma$ must also be real. Nonetheless, the suggestion is made that the results can be applied to a superconductor by replacing the skin depth ($\delta$) with the superconducting penetration depth ($\Lambda$--but symbolized herein by the more general term which is also applicable to non-superconductors, $\delta_A$). In making the London approximation, FKL assume the superconductor can be modeled by a purely imaginary conductivity. This also suggests that the wavelength inside the superconductor is infinite and that the real part of both the conductivity and intrinsic wave impedance is zero.[33] This simple London model may capture the bulk of the shielding behavior discussed in [26], but is at odds with the experimental finding of ac dissipation in superconductors.

Earlier work on transmissivity through superconducting films did allow the superconductor's conductivity to have a non-zero real part.[2-5] The transmissivity expression this work used[2-5] is derived in [4] and [5] and is the same result found by substituting a complex conductivity for the purely real conductivity in the transmissivity expression of [26]. Because no discussion is made of the wavelength inside the superconducting film in [2]-[5] or [26], one might question its relevancy to the transmissivity formula:

$$\frac{P_t}{P_{in}} = \frac{1}{(1+\sigma_1 d\eta_o/2)^2 + (\sigma_2 d\eta_o/2)^2} \quad (24)$$

We show that the wavelength inside the film is relevant to the application of Eq. 24 and that the difficulties discussed associated with Eq. 23b are exacerbated by a general complex conductivity. Because FKL do not discuss the intricacies of the averaging associated with Eq. 23a, one might imagine their quick derivation to admit a complex conductivity just as well as a purely real conductivity provided the restriction is changed from $d \ll \delta$, $\lambda_0$ to $d \ll \delta_A$, $\lambda_0$ where $\lambda_0$ is the free space wavelength. The only explicit restriction on the conductivity in [4] and [5] is that it obey the good conductor approximation. But Eq. 24 is applied elsewhere[40,44] and one wishes to know how Eq. 24 compares with Eq. 17 for arbitrary conductivities.

Figure 3 reveals the absolute value of the error in the transmissivity incurred by using Eq. 24 for a film suspended in free space with a complex resistivity of 0.07 - $i$4.5 $\mu\Omega$-cm as a function of the thickness of the film (normalized by $\delta_A$). With an excitation frequency of 1 PHz ($10^{15}$ Hz), this corresponds to $\delta_A \approx 0.31$ $\mu m$. For film thicknesses less than $\delta_A/100$ the error already approaches ~40%. At 1 PHz, $1/\omega\varepsilon \approx 1800$ $\mu\Omega$-cm and so the good conductor approximation is satisfied. Figure 4 evinces the identical plot as Fig. 3 except the complex conjugate of the resistivity of Fig. 3 has been used. Figure 4 reveals that Eq. 24 becomes applicable for much larger values of $d/\delta_A$ than for Fig. 3. In the case of Fig. 4 even for $d = \delta_A$ ($\approx 2.4$ $nm$), the error in the transmissivity is less than 40%.

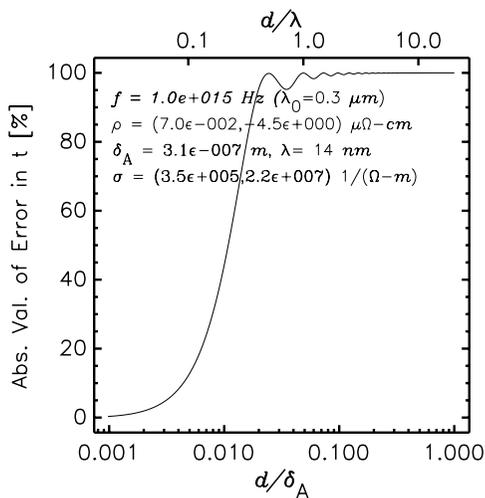

FIG. 3 Absolute value of the error in the transmissivity using Eq. 24 for a film suspended in free space with a complex resistivity of 0.07 - $i$4.5 $\mu\Omega$-cm as a function of the thickness of the film (normalized by $\delta_A$).

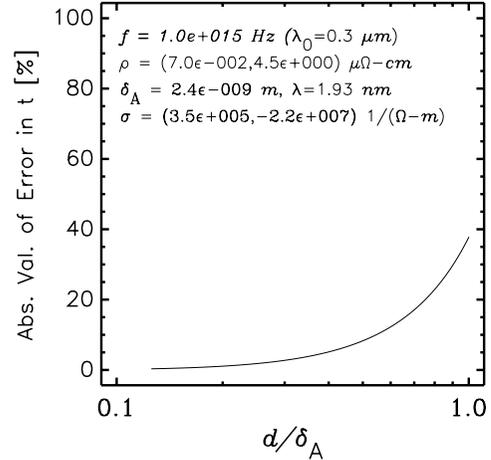

FIG. 4 Identical plot as Fig. 3, except the imaginary part of the resistivity has been negated and the thicknesses vary from 3 Å to $\delta_A$.

However the upper $x$-axis of Fig. 3 suggests that Eq. 24 does better when the additional constraint $d \ll \lambda$ is made. This is sensible in light of Eq. 23. In the case of Fig. 4, for $d \ll \delta_A \approx \lambda$ Eq. 24 fares well and our proposed restriction $d \ll \lambda$ survives two cases. Given a nominal lattice spacing of ~3Å, the values for $\delta_A$ and $\lambda$ in Fig. 4 approaches the continuum limit discussed at the end of sec. 2A.

Returning to the case of a resistivity given by 0.07 - $i$4.5 $\mu\Omega$-cm and described in Fig. 3, Fig. 5 manifests the corresponding field profile across the film from $x = 0$ to $x = d$. The two traces in Fig. 5 correspond to films with a thickness of $\delta_A/100$ and $\delta_A/50$ (depicted as a solid and dashed line respectively). The dotted line is a guide to the eye to illustrate the non-linear profile. As discussed in sec. 4A, this reveals the breakdown of the linear current profile assumed in Eq. 23b. Thus the applicability of Eq. 23b and/or Eq. 24 doesn't solely depend on $d/\delta_A$, but on the real and imaginary part of the conductivity in an unknown manner. By Eq. 13 the intrinsic impedance of whatever medium lies before the film plays a role in shaping the ratio of $E^+/E^-$ in Eq. 23b. The additional requirement $d \ll \lambda$ seems to impose a necessary restriction for the application of Eq. 24. For bulk superconductors well into the superconducting state, it is generally true that $-\Im(\sigma) \gg \Re(\sigma)$ ($> 0$) [43,44]. Figure 4 is an example where for a slab with characteristic superconducting parameters Eq. 24 is quite accurately correlated with a small $d/\delta_A$. Figure 2 is an example where for $\Im(\sigma) = 0$, $d \ll \delta, \lambda_0$ and $\sigma \gg \omega\varepsilon$ Eq. 24 again performs remarkably well. Dielectrics, semiconductors and semi-metals can have $\Im(\sigma) > 0$ and Fig. 3 is an example of such a material where Eq. 24 can incur errors exceeding 95% even though $d \ll \delta_A, \lambda_0$ and $|\sigma| \gg \omega\varepsilon$. Nonetheless with the additional constraint $d \ll \lambda$, Eq. 24 performs well in this example too.

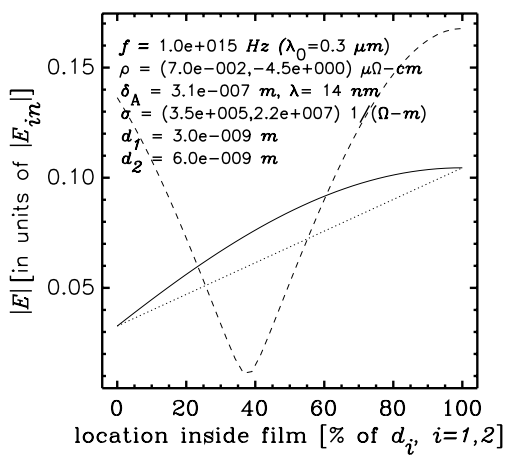

FIG. 5 Spatial profile of the magnitude of the $E$-field across a film with a conductivity identical to that of Fig. 3 having a thickness ($d$) less than $\delta_A/100$ and $\delta_A/50$ (depicted as a solid and dashed line respectively). The dotted line is a guide to the eye to illustrate the non-linear profile.

The formalism of sec. 2 and 3 dodges the need to compute the integral of the current density over the film in order to obtain the transfer function relations. In so doing we are able to use the formalism to exactly compute this integral. We close this subsection noting that none of the various restrictions ascribed to Eq. 24: $d \ll \delta_A, \lambda_0, \lambda$; $|\sigma| \gg \omega\varepsilon$, $-\Im(\sigma) \gg \Re(\sigma)$, or $\Im(\sigma) = 0$ apply to Eq. 17. Eq. 17 is an exact relation making only the continuum limit approximation and assuming planar isotropy. But these assumptions are also made in the derivation of Eq. 24.

### C. Multiple films

Beyond the appeal of the simple case of single suspended film entertained in secs. 4A and 4B, materials characterization requires the analysis of structures with many layers. Ceremuga-Mazierska [8] considers the transmission through a superconducting thin film-dielectric substrate structure. She adopts the steady-state picture for the transfer function involving the fields across the superconducting film in sec. 2.1, but then adopts the temporal series approach in deriving the transfer function for the fields across the substrate in sec. 2.2. Sec. 2.3 then combines these two transfer functions to produce a net transfer function. The problem with this mixing of these two approaches is that the variables have different meaning in the steady-state picture than in the temporal series picture. In sec. 2.1, the variable "$E_t$" is the total steady-state $E$-field at the back of the superconducting thin film. This total field is a superposition of the net forward propagating field with the net backward propagating field evaluated at the back of the superconducting thin film (or at the front of the dielectric substrate). This is a steady state field. In the temporal series picture in sec. 2.2 of [8], the transfer function does not relate this total $E$-field, but rather the initial temporal forward propagating field. This is not a steady-state field. Ref. [8] mistakenly equates the total steady-state field of sec. 2.1 to the initial forward propagating temporal field of sec. 2.2. The steady-state transfer function approach of secs. 2 and 3 herein offers both simplification and accuracy over the approach of [8].

Consider a four layered structure. The first layer is a low-loss dielectric with a relative permittivity of 9, a loss tangent of $10^{-5}$ and a thickness of 0.21 mm. Perhaps this is a passivation coating. The second layer is a superconducting film with a complex conductivity of $\sigma = (2-j500)*10^3$ $(\Omega\text{-}m)^{-1}$ and a thickness of 5 Å. The third layer consists of another low-loss dielectric with a relative permittivity of 16, a loss tangent of $10^{-7}$ and a variable thickness. The fourth layer is a bulk superconductor composed of the same conductivity as the film of layer 2.

Figure 6 reveals the power dissipation in each layer as a function of the incident frequency and the thickness of the third layer ($d_3$). The periodicity in $d_3$ is due to standing waves in the $3^{rd}$ film [23], [43]. The power dissipation is normalized by the incident power. Figs 6a-d display the dissipation in the low-loss overlayer, superconducting film, dielectric substrate and bulk superconductor respectively. Clearly one common feature is the strong dependence of the standing waves in the substrate on the behavior in each layer. As with the figures in [23], the maxima and minima in the various media occur at the same substrate thicknesses. This suggests that the fields throughout the structure are modulated by $d_3$ and that the reflected field is anti-correlated with the fields throughout the structure. Figure 7 depicts the total power dissipation in the four-layered structure of Fig. 6. As discussed in sec. 3A, this result can be achieved either by Eq. 15 or by summing the power dissipated in each layer. For each point in Fig. 7 the agreement between these two approaches is better than 2 parts in $10^7$ and is limited only by the precision of the computing machine.

Figure 6a reveals the fractional power dissipating in the low-loss overlayer. It is interesting to note that since the wave impedance for a low-loss dielectric can be approximated by $\eta_i \approx \eta_0 \cdot (1 + j\tan\delta_i/2)/\sqrt{\varepsilon_{ri}}$ where $\tan\delta_i$ is the loss tangent of medium $i$ [36], $\eta_i$ is roughly independent of frequency. Nevertheless Fig. 6a shows that the dissipative maxima increases with increasing frequency and the $d_3/\lambda_3$ location of the maxima also changes with frequency. This suggests there is a significant frequency dependence in $Z_1$ and/or $Z_2$. One also sees from Fig. 6a that for some $d_3$ values one can have less dissipation in the first film than for $d_3 = 0$.

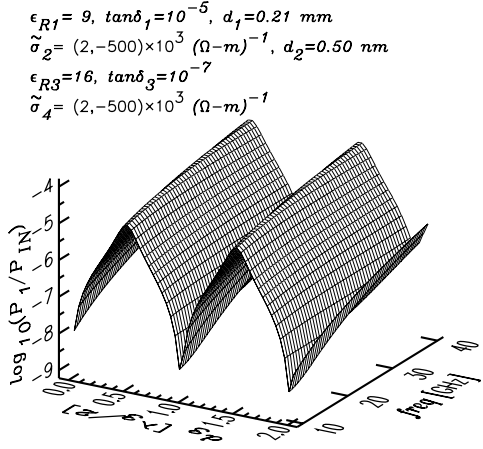

FIG. 6a Power dissipating in a low-loss overlayer, normalized by the incident power. As described in the text, the overlayer compromises the first layer in a 4-layer structure. The power dissipating in the overlayer is plotted as a function of the incident frequency and the thickness of the third layer, which is a low-loss dielectric (the thickness of which is shown in units of one-half the wavelength in the third medium).

Figure 6b is a case study in the richness of waves in

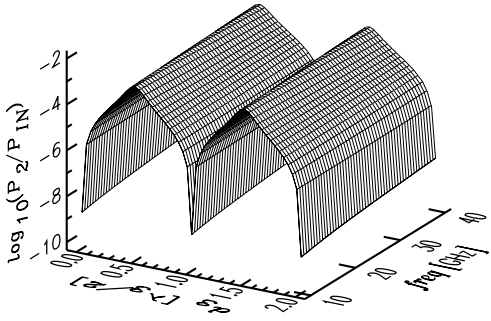

FIG. 6b Power dissipating in a superconducting film, normalized by the incident power. As described in the text, the superconducting film is the second layer in a 4-layer structure.

layered media. Changing the thickness of a low-loss dielectric by one quarter of a wavelength changes the power dissipation in the superconducting film by ~6 orders of magnitude! This is especially amazing when considering that the thickness of the superconducting film is only 5 Å. To appreciate this thickness, Fig. 8 shows $d_2$ to be well over 3 orders of magnitude less than

$\delta_{A2}$. Yet Fig. 6 indicates that for $d_3 \approx (2n+1)\lambda/4$ $n = 0,1,\ldots$ this seemingly insignificant film can dissipate more than any other layer!

Figure 6c depicts the dissipation in the third layer, a low-loss substrate having a relative permittivity of 16 and a loss tangent of $10^{-7}$. Apart from [23], the author is unaware of any other work which provides expressions that extricate the dissipation in a dielectric substrate despite the ubiquity of these substrates in ac measurements involving thin films. While the dissipation in the other layers are essentially identical for $d_3 = n\lambda/2$ for $n = 0,1,2 \ldots$; there is an exception to this symmetry for this layer. For $n = 0$, $d_3 = 0$ and so there cannot be any dissipation in a layer which has no thickness. For visualization purposes, we have arbitrarily given these $d_3 = 0$ data a dissipation value of $10^{-11}$.

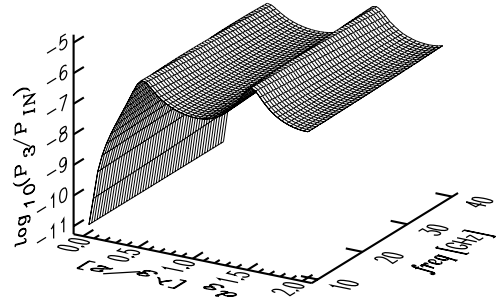

FIG. 6c Power dissipating in a dielectric substrate, normalized by the incident power. The dielectric substrate is the third layer in a 4-layer structure.

Figure 6d reveals a plot of Eq. 15. For a constant frequency, $\eta_4$ is constant because $\sigma_4$ is constant. The variation in the dissipation for constant frequency is then solely due to the changing magnitude of the tangential $H$-field at $x_4$. By Eq. 13d this also means that $|A_3|$ and $|B_3|$ change by the same factor.

Although figures in ref. [23] also display the dissipation in a superconducting film as a function of the thickness of an adjacent dielectric, there is a difference in the location of the peak dissipation in the film relative to the location of the dissipative minima in the film. Figures 2 and 3 of [23] place the dissipative maxima just before the integer one half wavelength thickness of the adjacent dielectric while in Fig. 6 this peak lies half way in between these thicknesses. Apparently this is related to the thickness of the superconducting film, because the same conductivity is used for the superconductors in [23] as herein. What *is* different is the thickness of the superconducting film which in [23] is 0.1 μm, while in Fig. 6 it is 5 Å.

To explore the significance of the dissipation in a 5 Å thick film, one is interested in knowing $\delta_A$ for that film. Figure 8 depicts $\delta_{A2}$ as a function of frequency and one sees that $\delta_{A2}$ is many orders of magnitude larger than $d_2$. Figure 9 resolves this difference in the location of the maxima showing the evolution of the peak between these two locations, as a function of the superconducting film's thickness. For definiteness, the frequency is fixed at 10 GHz. At this frequency, $\delta_{A2}$ is ~5 $\mu m$ so that for the range of $d_2$ shown, $d_2$ is always less than 1/5$^{th}$ of $\delta_{A2}$. Nonetheless variation in this lower quintile allows the peak to migrate from a location very close to $d_3 = n\lambda/2$ to one very close to $(2n+1)\lambda/4$. The location of this peak can be understood by the arguments presented with regard to the recently discovered superconducting length scale $\delta_0$ [45]. Although the study in [45] involves a superconducting film with a fixed thickness, the argument for qualitatively different behavior for $\delta > \delta_0$ or $\delta < \delta_0$ applies to a film of *any* thickness. Undoubtedly an analogous argument involving impedance mismatch explains the location of the maxima in Fig. 9.

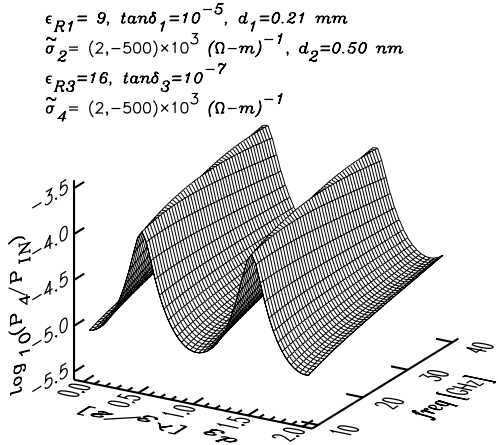

FIG. 6d Power dissipating in a bulk superconductor, normalized by the incident power. The superconductor constitutes the fourth layer in a 4-layer structure.

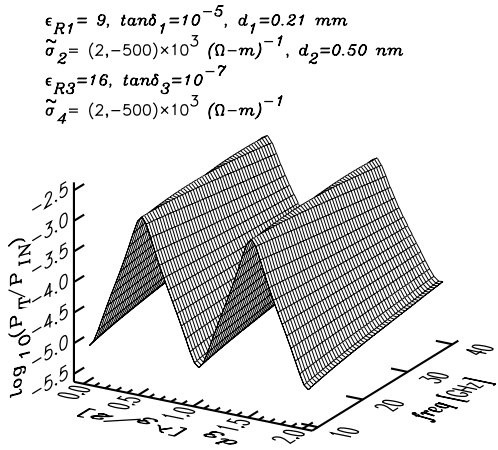

FIG. 7 Total power dissipation of all four layers in Fig. 6.

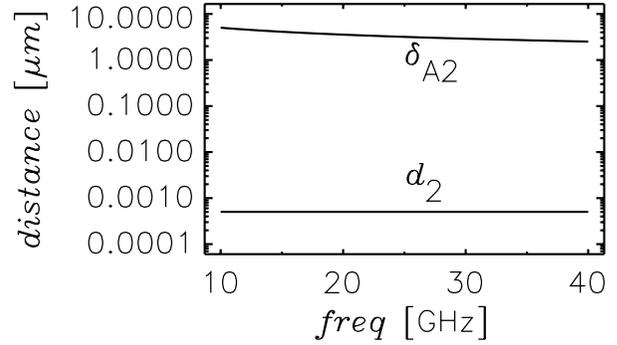

FIG. 8 Amplitude attenuation length scale ($\delta_{A2}$) of the film of Fig. 6b as a function of frequency. Also superimposed is the thickness of this second film, $d_2$.

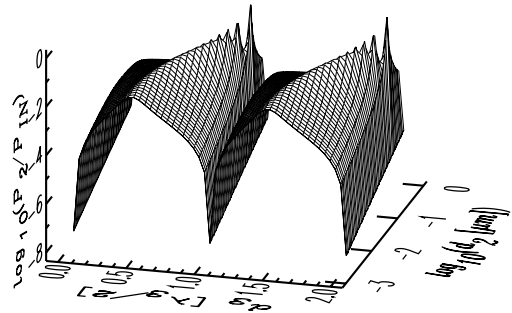

FIG. 9 Power dissipation in the superconducting thin film, as a function of the film's thickness and the thickness of an adjacent dielectric. The frequency is held constant and the 4-layered structure is identical to that of Fig. 6, except the thickness of the superconducting film instead of the frequency is varied.

Figure 9 also reveals why the averaging approach of Tinkham *et al.* discussed in sec. 1 is problematic. For $d_2 \ll \delta_{A2}$ the area under $P_3$--for a range of frequencies--depends on the thickness of the substrate.

In the nomenclature of Fig. 6, Klein *et al.*[37] claim that for $d_3$ very different from $n\lambda_3/2$ ($n=1,2,3,...$) the dissipation in Figs. 6c and 6d should be "suppressed significantly." However

Figs. 6, 7 and 9 show that for such $d_3$'s, these transmission losses can be maximized.

Other details can be found. The frequency independence of the periodicity in the dissipative minima of Fig. 6 can be understood by the same arguments used in the derivation of the length scale $\Delta d$ in [43]. As discussed in association with the text of Eq. 6a in [43], the phase of the reflection coefficients on either side of the film does not change as a function of the thickness of the film so that the periodicity of the film thickness is precisely $\lambda/2$. This is also confirmed by both Fig. 6 and Fig. 10, which displays the same parameter as Fig. 6b but represents the case where the first layer is removed.

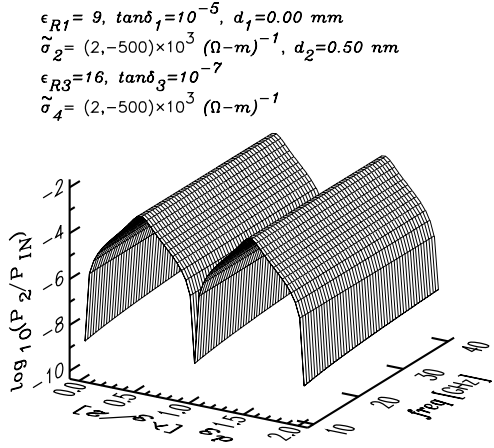

FIG. 10 Same as Fig. 6b except the first layer is removed. The maxima no longer exhibit the significant frequency dependence as shown in Fig. 6b.

In contrast to Fig. 6, Fig. 10 reveals that the maxima no longer exhibits the significant frequency dependence shown in Fig. 6. The frequency dependence in the dissipative maxima of Fig. 10 is due to the changing phase of the reflection coefficient $A_3/B_3$ as discussed in [43].

In the case of Fig. 10, the picture of Eq. 6b of [43] explains the independence of the dissipation maxima with frequency. In this picture, changing the frequency or conductivity can move the system toward or away from a dissipative maxima. The explanation is that the phase of the left or backward reflection coefficient depends on frequency for the case of Fig. 6, but is relatively insensitive for the case of Fig. 10. The backward reflection coefficient is determined by the ratio of the impedance looking backward at $x_3$ ($Z_{back}$) with $\eta_3$. Figure 11 confirms this explanation as it plots the phase of the backward looking impedance as a function of frequency for the cases involving Fig. 6 and Fig. 10. Since $\eta_3$ is relatively independent of frequency, the frequency dependence of this backward reflection coefficient is dominated by the frequency dependence of this backward looking impedance. Since the forward looking impedance at $x_4$, $Z_4$, is the same for the case of Fig. 6 and Fig. 10, it cannot explain this disparity in frequency sensitivity. The backward looking impedance can be computed from its forward-looking counterpart impedance determined by inverting the $x$-axis so that we consider medium 1 to be a superconductor with a complex conductivity of $\sigma_1 = (2-j500)*10^3$ $(\Omega\text{-}m)^{-1}$ and a thickness of $d_1 = 5$ Å, medium 2 to be a low-loss dielectric with a relative permittivity of 9, a loss tangent of $10^{-5}$ and a thickness of 0.21 mm and finally the third medium is semi-infinite free space. The phase of this equivalent forward-looking impedance as a function of frequency is shown in Fig. 11. The solid curve represents the case with the second medium present and the dashed curve represents the case with the second medium absent. Consistent with the picture of Eq. 6b of [43], Fig. 11 shows the phase of the impedance with the second medium present to be much more sensitive to frequency than without it.

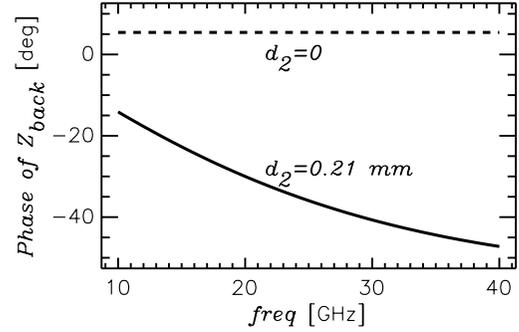

FIG. 11 Frequency dependence of the phase of the impedance looking backward (or to the left) from $x_3$ of Fig. 6 with (solid) and without (dashed) the first film.

## 5. CONCLUSION

The analytic tools for multilayer structures suitable for material science have progressed over time. Tinkham[2] has provided a transmissivity expression that is very simple and reasonably accurate over some unknown range in parameter space ($\sigma$, $d$ and $n$ where $n$ is the index of refraction of the substrate). By exactly accounting for the thickness of the substrate, Ceremuga-Mazierska[8] improves upon the Tinkham relation and is better poised to account for the effects of the standing waves in the substrate, which as shown in Fig. 6, can dominate the behavior in each layer. However Ceremuga-Mazierska confuses a steady-state total field for an initial instantaneous temporal forward propagating field and thereby incurs an error.

Fahy, Kittel and Louie (FKL)[26] achieve simplification in developing expressions for the transmissivity, reflectivity and absorptivity of a film suspended in free space. They assume a linear current density profile vs. position in the film, which we have shown to be non-linear under some conditions. Although their expressions are verisimiltudinous over a large range in thickness and conductivity they incur errors of up to 100% for very thin films.

These various approaches[2-5,8,26] seeking simplification to a steady-state problem have emerged because there has not been a steady state formalism that can intuitively and simply answer such questions. The temporal series approach--which its infinite series and profusion of terms--is a difficult and painful means to solve multilayered steady-state problems.

Even when the temporal series approach was not used to solve steady-state problems, the steady-state approach that has been used avoided using the surface wave impedance ($Z_{sw}$ or $Z_i$) and thereby incurred its own profusion of terms.[11,15] By using $Z_i$ at each surface we have developed a general formalism that is intuitive and exact and that avoids the profusions of terms that have besieged these other methods. By avoiding either method which leads to a profusion of terms, we have obtained a simple and intuitive formalism. Because the complexity of this formalism is couched in terms involving the ratio of the surface wave impedance with its own or an adjacent intrinsic impedance (*i.e.*, $Z_i/\eta_j$, where $j = i$ or $i-1$), simplification of the overall transfer function immediately follows simplification of the term $Z_i/\eta_j$. Unlike the uncertainty surrounding the applicability of the Tinkham transmissivity formula[2-5] or the FKL expressions[26] which do not have a well-defined application, the conditions limiting application of the simplified transfer function is limited only by those same conditions that ushered simplification in $Z_i/\eta_j$. For example the simplification in $Z_i$ determined for bimetallic structures[32] can be applied to usher a host of simplified transfer functions concerning bimetallic structures. These simplified transfer functions then have their application limited by the same conditions which confine the various $Z_i$ approximations.

Stated formally, provided the length scales being considered are sufficiently large so that charge separation is negligible and the material is linear in both Ohm's law and in the constitutive equations, Maxwell's equations yield the Helmholtz equation. Under these general conditions, if we further limit ourselves to uniaxial stratified media subjected to normal radiation and evaluated in the steady-state (Fig. 1), then we have developed a formalism which explicitly describes the fields in each layer.

This formalism determines the fields on the basis of two relationships. The first recursively determines the surface wave impedance ($Z_{sw}$) at each surface ($Z_i$) starting at the back surface and sequentially moving to the front surface. The second relationship converts a generic boundary condition relation to the particular boundary condition relation by substituting the particular $Z_i$ for the generic $Z_{sw}$. Having determined $Z_i$ at each interface, one can then relate the incident field to the steady-state field in the first layer, the second layer, and so on by cascading the corresponding transfer functions.

With the incident radiation originating from a loss-less medium [34], we provide numerically robust algorithms for computing the steady-state power dissipation in each layer. One algorithm invokes the good conductor approximation (Eq. 16) while the other is exact (Eq. 20).

This work provides an intuitive and physical formalism for the fields and the power dissipated in multilayered structures and obtains exact results. We use this work to correct other work involving multilayer electrodynamics[2-5,8,37] and boundary conditions[39,40]; to provide scope to a popular transmissivity expression (Eq. 24)[2-5,26,40,44] and to correct putative beliefs about the linearity of the current density profile in thin films[8,26] by showing a highly non-linear current density profile.

We show via an example of a four-layered structure, the dependence of the dissipation in each of the layers on the thickness of a particular layer. The richness of the physics possibilities in waves in layered media is manifest through this sample system whereby changing the thickness of this low-loss dielectric by one-quarter of a wavelength changes the power dissipation in a 5 Å superconducting film by ~6 orders of magnitude! Although its thickness is well over 3 orders of magnitude less than $\delta_A$--this superconducting film dissipates more energy than any other layer!

Finally although our starting point--the Helmholtz equation--emerged from electrodynamics, this work is generalizable to other systems involving the steady-state response of waves in layered media. Such systems include acoustical and quantum mechanical systems.

## 6. ACKNOWLEDGEMENTS

The author thanks Keith Bowen for discussions on the significance of the recursive treatment in [17] and David Stroud for noting an error in a previous version of this manuscript.

* pbeeli@uh.edu